\documentclass[twocolumn,superscriptaddress,showpacs,oneside,prl]{revtex4}%
\usepackage{amsmath}
\usepackage{amssymb}
\usepackage{amsfonts}
\usepackage{graphicx}%
\begin{document}
\title{Experimental Quantum Cloning with Prior Partial Information}
\author{Jiangfeng Du}
\email{djf@ustc.edu.cn}
\affiliation{Hefei National Laboratory for Physical Sciences at Microscale \& Department
of Modern Physics, University of Science and Technology of China, Hefei,
Anhui 230026, PR China}
\affiliation{Department of Physics, National University of Singapore, 2 Science Drive 3,
Singapore 117542}
\affiliation{Centre for Quantum Computation, DAMTP, University of Cambridge, Wilberforce
Road, Cambridge CB3 0WA U.K.}
\author{Thomas Durt}
\affiliation{TONA-TENA Free University of Brussels, Pleinlaan 2, B-1050 Brussels, Belgium.}
\author{Ping Zou}
\affiliation{Hefei National Laboratory for Physical Sciences at Microscale \& Department
of Modern Physics, University of Science and Technology of China, Hefei,
Anhui 230026, PR China}
\author{Hui Li}
\affiliation{Hefei National Laboratory for Physical Sciences at Microscale \& Department
of Modern Physics, University of Science and Technology of China, Hefei,
Anhui 230026, PR China}
\author{L.C. Kwek}
\affiliation{Department of Natural Sciences, National Institute of Education, Nanyang
Technological University, 1 Nanyang Walk, Singapore 637616}
\author{C.H. Lai}
\affiliation{Department of Physics, National University of Singapore, 2 Science Drive 3,
Singapore 117542}
\author{C.H. Oh}
\affiliation{Department of Physics, National University of Singapore, 2 Science Drive 3,
Singapore 117542}
\author{Artur Ekert}
\affiliation{Centre for Quantum Computation, DAMTP, University of Cambridge, Wilberforce
Road, Cambridge CB3 0WA U.K.}
\affiliation{Department of Physics, National University of Singapore, 2 Science Drive 3,
Singapore 117542}

\begin{abstract}
When prior partial information about a state to be cloned is available, it can
be cloned with a fidelity higher than that of universal quantum cloning. We
experimentally verify this intriguing relationship between the cloning
fidelity and the prior information by reporting the first experimental optimal
quantum state-dependent cloning, using nuclear magnetic resonance techniques.
Our experiments may further have important implications into many quantum
information processing protocols.

\end{abstract}
\pacs{03.67.Dd, 76.60.-k}
\maketitle

One of the most striking and important implication of quantum mechanical laws
to the information theory is that an unknown quantum state \textit{cannot} be
copied exactly, as observed by Wootters and Zurek \cite{1}, and Ghirardi and
Weber \cite{2}. However if it is not demanded to copy the state exactly, then
it is possible to clone it \textit{approximately} \cite{2-2}. Since Buzek and
Hillery proposed their \textit{universal quantum cloning machine} (UQCM)
\cite{3}, lots of efforts had been made to the theoretical investigation of
this kind of cloner \cite{4}. Also a number of experimental implementations of
$1\rightarrow2$ qubit universal cloning had been proposed \cite{5}.

A universal cloning machine for qubits produces copies of equal quality for
all possible input states, with a fidelity being $5/6\approx0.833$ for a
$1\rightarrow2$ optimal UQCM. However, if partial prior knowledge of the state
to be copied is available, or it is known to belong to a subset of all
possible input states, then it is possible to clone the state with a higher
fidelity. For instance this occurs in the case of the quantum phase covariant
cloning machine (QPCCM) \cite{6,6-2,9,8}. A QPCCM clones a equatorial qubit
state, which is in the form of $|\psi\rangle=(|0\rangle+e^{i\phi}%
|1\rangle)/\sqrt{2}$, with an optimal fidelity being $0.854$, higher than that
of a UQCM.

In a previous paper \cite{10}, we have shown why two qubits are enough for
$1\rightarrow2$ optimal phase-covariant cloning in agreement with
Niu-Griffiths scheme \cite{6-2}. In the present letter, we constructed the
two-qubit quantum logic circuit for the optimal quantum state-dependent
cloning, and experimentally observe the interesting relationship between the
cloning fidelity and the prior information available. In our case, for a qubit
state of the form%
\begin{equation}
|\psi\rangle=\cos\frac{\theta}{2}|0\rangle+e^{i\phi}\sin\frac{\theta}%
{2}|1\rangle,\label{eq 1}%
\end{equation}
the prior knowledge is $\theta\in\lbrack0,\pi]$, the polar angle of the
corresponding vector in the Bloch sphere, while the azimuthal angle is
completely unknown, i.e. uniformly distributed between $[0,2\pi]$. When
$\theta$ is fixed to be $\pi/2$, our quantum state-dependent cloning reduced
to the conventional quantum optimal phase-covariant cloning. Our cloning, as
will be presented in the following, indeed achieves the optimal fidelities for
any given polar angle $\theta$, which is presented in Ref. \cite{8}.

We experimentally implement the $1\rightarrow2$ cloning of qubit states with
\textit{a priori} known value of the polar angle $\theta$. The observed
fidelities are higher than that of the UQCM, with a minimal fidelity being
exactly that of the QPCCM, which is, however, still higher than that of the
UQCM. The experimental observation agrees well with theoretical predictions,
showing the interesting relationship between the cloning fidelities and the
prior information of the input state. As quantum cloning is intimately related
to many quantum information processing protocols, our experimental observation
may further have important implications into them.

\begin{figure}[t]
\begin{center}
\includegraphics[width=0.85\columnwidth]{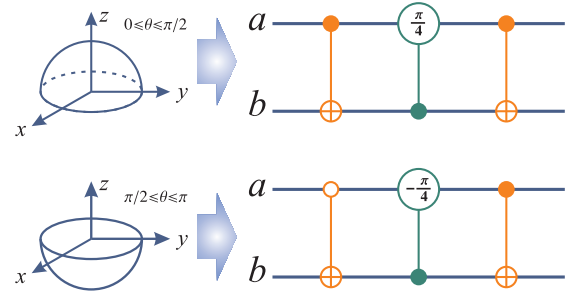}
\end{center}
\caption{The quantum logic circuit for our optimal quantum state-dependent
cloning machine. Qubit $a$ is the one to be cloned, initially in the state
$|\psi\rangle$ in Eq. (\ref{eq 1}). While qubit $b$ is the blank one which is
initially in the state $|0\rangle$. For states in the northern hemisphere, the
upper circuit is used, while for states in the southern hemisphere the lower
circuit is adopted.}%
\label{logicgate}%
\end{figure}

The quantum logic circuits for our optimal quantum state-dependent cloning is
described in Fig. \ref{logicgate}. We use two different circuits for the case
where the state resides in the northern and the southern hemisphere,
respectively. Qubit $a$ contains the state to be cloned as in Eq.
(\ref{eq 1}), while qubit $b$ is the blank one initially in the state
$|0\rangle$.

The unitary operator representing the quantum logic circuit for the northern
hemisphere, i.e. for $\theta\in\lbrack0,\pi/2]$, is{}%
\begin{align}
U_{n} &  =|00\rangle\langle00|+|11\rangle\langle11|+\nonumber\\
&  \frac{1}{\sqrt{2}}\left(  |01\rangle\langle01|+|10\rangle\langle
10|+|01\rangle\langle10|-|10\rangle\langle01|\right)  .\label{eq un}%
\end{align}
While for the southern hemisphere, i.e. $\theta\in\lbrack\pi/2,\pi]$, the
unitary operator representing the corresponding circuit is%
\begin{align}
U_{s} &  =|00\rangle\langle01|+|11\rangle\langle10|+\nonumber\\
&  \frac{1}{\sqrt{2}}\left(  |01\rangle\langle00|+|10\rangle\langle
00|+|10\rangle\langle11|-|01\rangle\langle11|\right)  .\label{eq us}%
\end{align}
With the initial state being $|\psi\rangle$ in Eq. (\ref{eq 1}) for qubit $a$
and $|0\rangle$ for qubit $b$, the two identical clones of $|\psi\rangle$ can
be easily verified to be described by density matrices%
\begin{equation}
\rho_{a}=\rho_{b}=\left(
\begin{array}
[c]{cc}%
\cos^{2}\dfrac{\theta}{2}+\dfrac{1}{2}\sin^{2}\dfrac{\theta}{2} & \dfrac
{1}{2\sqrt{2}}e^{-i\phi}\sin\theta\\
\dfrac{1}{2\sqrt{2}}e^{i\phi}\sin\theta & \dfrac{1}{2}\sin^{2}\dfrac{\theta
}{2}%
\end{array}
\right) \label{eq clone-n}%
\end{equation}
for $\theta\in\lbrack0,\pi/2]$, while%
\begin{equation}
\rho_{a}=\rho_{b}=\left(
\begin{array}
[c]{cc}%
\dfrac{1}{2}\cos^{2}\dfrac{\theta}{2} & \dfrac{1}{2\sqrt{2}}e^{-i\phi}%
\sin\theta\\
\dfrac{1}{2\sqrt{2}}e^{i\phi}\sin\theta & \dfrac{1}{2}\cos^{2}\dfrac{\theta
}{2}+\sin^{2}\dfrac{\theta}{2}%
\end{array}
\right) \label{eq clone-s}%
\end{equation}
for $\theta\in\lbrack\pi/2,\pi]$. It then immediately follows that the
fidelity of our quantum state-dependent cloning is%
\begin{align}
F(\theta) &  =\operatorname*{Tr}(\rho_{a}|\psi\rangle\langle\psi
|)=\operatorname*{Tr}(\rho_{b}|\psi\rangle\langle\psi|)\nonumber\\
&  =\left\{
\begin{array}
[c]{ll}%
\dfrac{1}{2}\sin^{2}\dfrac{\theta}{2}+\cos^{4}\dfrac{\theta}{2}+\dfrac
{\sqrt{2}}{4}\sin^{2}\theta, & \theta\in\left[  0,\dfrac{\pi}{2}\right] \\
\dfrac{1}{2}\cos^{2}\dfrac{\theta}{2}+\sin^{4}\dfrac{\theta}{2}+\dfrac
{\sqrt{2}}{4}\sin^{2}\theta, & \theta\in\left[  \dfrac{\pi}{2},\pi\right]
\end{array}
\right.  ,\label{eq fidelity}%
\end{align}
which has already been expressed as a function of the polar angle $\theta$,
for all $\theta\in\lbrack0,\pi]$. This fidelity is exactly the optimal cloning
fidelity which could be achieved for our state-dependent cloning, as proved in
Ref. \cite{8}. It would also be interesting to note that the cloning fidelity
is independent on the azimuthal angle $\phi$, which implies that the cloning
is equally efficient for all the states with the same value of polar angle
$\theta$. In our case, it is interesting to investigate the relationship
between the fidelity $F(\theta)$ and the corresponding \textit{entropy} of the
input state (a state with fixed known polar angle $\theta$ but completely
unknown azimuthal angle $\phi$). Here, for states $|\psi\rangle$ in Eq.
\ref{eq 1} with fixed $\theta$ but unknown $\phi$, the \textit{average} state
can be represented by a density matrix $\varrho=\frac{1}{2\pi}\int|\psi
\rangle\langle\psi|d\phi$, and its uncertainty can be expressed by the
entropy, i.e. $S(\varrho)=-p\log_{2}p-(1-p)\log_{2}(1-p)$ where $p$ and $1-p$
are eigenvalues of $\varrho$. In terms of the polar angle $\theta$, the
entropy can then be expressed as a function of $\theta$:%
\begin{equation}
S(\theta)=-\cos^{2}\frac{\theta}{2}\log_{2}\cos^{2}\frac{\theta}{2}-\sin
^{2}\frac{\theta}{2}\log_{2}\sin^{2}\frac{\theta}{2}.\label{eq entropy}%
\end{equation}

\begin{figure}[b]
\begin{center}
\includegraphics[width=\columnwidth]{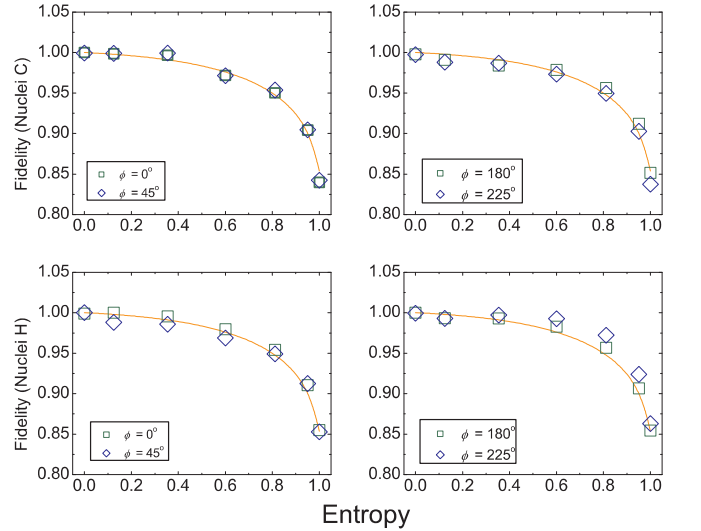}
\end{center}
\caption{The experimental fidelity versus the entropy of the input state. The
theoretical values of fidelities are plotted as solid lines. The upper two
figures are corresponding to the experimental result of qubit $a$ with
different azimuthal angles while the lower two figures are corresponding to
that of qubit $b$.}%
\label{data}%
\end{figure}

In our nuclear magnetic resonance (NMR) experimental implementation of the
above quantum state-dependent cloning, we use $^{13}$C nucleus and the $^{1}$H
nucleus in Carbon-13 labeled chloroform (Cambridge Isotopes) dissolved in
d$_{6}$ acetone as qubit $a$ and qubit $b$ correspondingly. Experimentally,
the reduced Hamiltonian of the two-spin ensemble is given by $H=\omega
_{a}I_{z}^{a}+\omega_{b}I_{z}^{b}+2\pi JI_{z}^{a}I_{z}^{b}$, where the first
two terms describe the free procession of spin $a$ ($^{13}$C) and spin $b$
($^{1}$H) around the static magnetic field (along direction $\widehat{z}$)
with frequencies $100$MHz and $400$MHz, respectively. $I_{z}^{a(b)}=\frac
{1}{2}\sigma_{z}^{a(b)}$ is the angular moment operator of $a$ ($b$) in
direction $\widehat{z}$, and the third term is the $J$ coupling of the two
spins with $J=214.5$Hz. $^{13}$C nucleus's $T_{1}$\ relaxation time is $17.2$s
and its $T_{2}$ relaxation time is $0.35$s. $^{1}$H nucleus's $T_{1}$
relaxation time is $4.8$s and it's $T_{2}$ relaxation time is $3.3$s. The
experiments must be implemented within the relaxation time before the quantum
coherence is lost due to the unavoidable decoherence.

We prepare the pseudo-pure state $|00\rangle$ following the same proposal as
presented in Ref. \cite{7}. Then, by rotating qubit $a$, we get the two-qubit
initial state as $(\cos\theta|0\rangle+e^{i\phi}\sin\theta|1\rangle)/\sqrt
{2}\otimes|0\rangle$. In the experiments, we set $\theta=n\cdot\pi/12$, which
determines the entropy of the input state, for $n=0,1,\cdots,12$ and
$\phi=m\cdot\pi/4$ for $m=0,1,\cdots,7$. For each set of $(\theta,\phi)$ the
full process of the cloning in Fig. \ref{logicgate} is implemented and the
fidelity is extracted. The NMR pulse sequences for our quantum state-dependent
cloning are developed by replacing unitary operations with idealized sequences
of NMR pulses and delays. The resulting sequences are further simplified by
appropriately combining radio-frequency (rf) pulses when convenient. All the
rf pulses on a single qubit are hard pulses which give nearly no affection on
the state of the other qubit due to the heteronuclear sample we used.

The specific pulse sequences for preparing the pseudo-pure state $|00\rangle$
is%
\begin{equation}
R_{x}^{b}(\pi/3)-G_{z}-R_{x}^{b}(\pi/4)-\tau_{1}-R_{-y}^{b}(\pi/4)-G_{z}%
,\label{eq ps-state}%
\end{equation}
which is read from left to right, radio-frequency pulses are indicated by
$R_{axis}^{spins}\left(  angle\right)  $, and are applied to the spins in the
superscript, along the axis in the subscript, by the angle in the brackets.
For example, $R_{x}^{b}(\pi/3)=e^{-i(\pi/3)\widehat{\sigma}_{x}/2}$ denotes
$\pi/3$ selective pulse that acts on the qubit $b$ about $\widehat{x}$, and so
forth. $G_{z}$ is the pulsed field gradient along the $\widehat{z}$ axis to
annihilate transverse magnetizations, dashes are for readability only, and
$\tau_{1}$ represents a time interval of $1/(2J)$, corresponding to an
evolution operation as $e^{-i\frac{\pi\tau_{1}J}{2}\widehat{\sigma}_{z}%
^{a}\widehat{\sigma}_{z}^{b}}$.

The NMR pulse sequence for the quantum logic circuit of the northern
hemispheres is\begin{widetext}%

\begin{equation}
R_{-y}^{b}(\pi/2)-\tau_{1}-R_{-x}^{b}(\pi/2)-R_{y}^{a}(\pi/2)-\tau_{2}%
-R_{-y}^{a}(\pi/2)-R_{x}^{a}(\pi/4)-R_{x}^{b}(\pi/2)-\tau_{1}-R_{-y}^{b}%
(\pi/2)\label{eq ps-n}%
\end{equation}
and for the southern hemispheres is%
\begin{equation}
R_{y}^{b}(\pi/2)-\tau_{1}-R_{-x}^{b}(\pi/2)-R_{-y}^{a}(\pi/2)-\tau_{2}%
-R_{y}^{a}(\pi/2)-R_{-x}^{a}(\pi/4)-R_{x}^{b}(\pi/2)-\tau_{1}-R_{-y}^{b}%
(\pi/2)\label{eq ps-s}%
\end{equation}
\end{widetext}with the same meaning of notions as in the sequence in Eq.
(\ref{eq ps-state}) but $\tau_{2}$ representing a time interval of $1/(4J)$.
Alert readers may argue that the operation realized by the pulse sequences as
in Eqs. (\ref{eq ps-n}, \ref{eq ps-s}) are actually different from Eqs.
(\ref{eq un}, \ref{eq us}) by certain rotations around the $z$-axis which
should be added before we implement the pulse sequences. Specifically,
$R_{z}^{a}\left(  \pi\right)  \otimes R_{z}^{b}\left(  \pi\right)  $ should be
added before the sequence in Eq. (\ref{eq ps-n}) is implemented, while
$R_{z}^{a}\left(  \pi\right)  $ before that in Eq. (\ref{eq ps-s}). However,
those rotations can always be absorbed into the abstract reference frames in
NMR experiments, hence no actual rotations are required in our case.

In the language of Bloch sphere, the state of a single qubit can be
represented by a density matrix of the form $\rho=\frac{1}{2}(\widehat{\sigma
}_{0}+\widehat{r}\cdot\widehat{\sigma})$, where $\widehat{\sigma}_{0}$ is the
identity operator, $\widehat{\sigma}_{\mu}$ ($\mu=x,y,z$) are the usual Pauli
matrices, and $\widehat{r}=(r_{x},r_{x},r_{x})$ is a real vector, which is of
length less than one for mixed states while equal to one for pure states. Let
$\rho_{0}=|\psi\rangle\langle\psi|=\frac{1}{2}(\widehat{\sigma}_{0}%
+\widehat{r}_{0}\cdot\widehat{\sigma})$ be the density matrix for the initial
state $|\psi\rangle$ in Eq. (\ref{eq 1}) while $\rho=\rho_{a}=\rho_{b}%
=\frac{1}{2}(\widehat{\sigma}_{0}+\widehat{r}\cdot\widehat{\sigma})$ for the
copy. The fidelities of the cloning in Eq. (\ref{eq fidelity}) then can be
expressed as $F(\theta)=\operatorname*{Tr}(\rho_{0}\cdot\rho)=\frac{1}%
{2}(1+\widehat{r}_{0}\cdot\widehat{r})$. To experimentally determine the
fidelities of the cloning, we need to measure the $x$, $y$, $z$ part of
$\widehat{r}$. This can be accomplished by phase sensitive detectors for
measuring the $x$ and $y$ parts, while by adding a gradient pulse and a
$\pi/2$ pulse for measuring the $z$ parts of the copies. And then the
fidelities can be obtained directly from the experimental data.

All our experiments are conducted at room temperature and normal pressure on a
Bruker AV-400 spectrometer. The experimental fidelities $F(\theta)$ are
presented in Fig. \ref{data}, as a function of the entropy $S(\theta)$ of the
input state. We observe that as the prior information increases, i.e. the
entropy decreases, the fidelity monotonously increases. Our experimental
results coincides with the theoretical predictions. The small errors are
mainly due to imperfect pulses, the variability over time of the measurement
process, and inhomogeneity of magnetic field.

It is worth noting that the fidelity of our state-dependent cloning are higher
than either that of the UQCM or of the QPCCM, while with a minimal value of
fidelity being exactly that of the QPCCM. Furthermore, when $\theta$
approaches $0$ or $\pi$, the corresponding fidelity increases. This is an
interesting observation from an informational prospective. Indeed, when
$\theta$ approaches $0$ or $\pi$, there are less states in the Bloch sphere
with the same polar angle of $\theta$. Hence there is less uncertainty of the
input state, i.e. more prior information is known about the input state, and
consequently higher fidelity is attainable in cloning this subset of states.
Moreover, in our present letter, the optimal cloning happens to be exactly the
square root of a SWAP operation. This is an interesting observation when
taking into account the interesting connection between quantum cloning and
quantum communication. Roughly speaking, this is because a SWAP permutes Bob's
state and Eve's state, which gives all the information to Eve and nothing to
Bob, while not doing anything sends all the information to Bob and nothing to
Eve: optimal cloning is in-between: the same information for Bob and for Eve.

In conclusion, we experimentally implement the $1\rightarrow2$ quantum optimal
state-dependent cloning for qubits. Our experimental observation highlights
the interesting relationship between the optimal cloning fidelities and the
prior information of the input states. The more is the prior information
available for the input states, the higher optimal cloning fidelity is
attainable. We further expect our experimental observation have important
implications into many quantum information processing protocols.

We thank N. Cerf and Dr. H. Fan for helpful discussions. This project was
supported by the National Nature Science Foundation of China (Grants. No.
10075041 and No. 10075044) and Funded by the National Fundamental Research
Program (2001CB309300). We also thank supports from the ASTAR Grant No.
012-104-0040 and Temasek Project in Quantum Information Technology (Grant No.
R-144-000-071-305). T.D. thanks supports from the Flemish Fund for Scientific
Research, the Inter-University Attraction Pole Program of the Belgian
government under grant V-18, the Concerted Research Action Photonics in
Computing, and the research council (OZR) of the VUB.


\begin{thebibliography}{99}                                                                                               %
\bibitem {1}W.K. Wootters and W.H. Zurek, Nature (London) \textbf{299}, 802 (1982).

\bibitem {2}G.C. Ghirardi and T. Weber, Nuovo Cimento Soc. Ital. Fis., B
\textbf{78}, 9 (1983).

\bibitem {2-2}D. Dieks, Phys. Lett. A 92, 271 (1982).

\bibitem {3}V. Buzek and M. Hillery, \pra \textbf{54}, 1844 (1996).

\bibitem {4}N. Gisin and S. Massar, \prl \textbf{79}, 2153 (1997); D. Bruss,
A. Ekert, and C. Marcchiavello, \prl \textbf{81}, 2598 (1998); L.M. Duan and
G.C. Guo, \prl \textbf{80}, 4999 (1998); C. Simon, G. Weihs, and A. Zeilinger,
\prl \textbf{84}, 2993 (2000); V. Buzek and M. Hillery, \prl   \textbf{81},
5003 (1998); S.L. Braunstern et al., \prl \textbf{86}, 4938 (2001); G.M.
D'Ariano, F. De Martini, M.F. Sacchi, \prl \textbf{86}, 914 (2001).

\bibitem {5}H.K. Cummins et al., \prl \textbf{88}, 187901 (2002); A.
Lamas-Linares et al., Science \textbf{296}, 712 (2002); S. Fasel et al., \prl
\textbf{89}, 107901 (2002); Y.F. Huang, \pra \textbf{64}, 012315 (2001).

\bibitem {6}C.A. Fuchs et al., \pra\textbf{56}, 1163 (1997); R.B. Griffiths
and C.S. Niu, \pra\textbf{56}, 1173 (1997); D. Bruss et al., \pra\textbf{62}
012302 (2000).

\bibitem {6-2}C.S. Niu and R.B. Griffiths, \pra\textbf{60}, 2764 (1999).

\bibitem {9}V. Karimipour and A.T. Rezakhani, \pra\textbf{66}, 052111 (2002).

\bibitem {8}J. Fiur\'{a}\v{s}ek, \pra\textbf{67}, 052314 (2003).

\bibitem {10}T. Durt and J. Du, quant-ph/0309072 (To appear in \pra).

\bibitem {7}J. Du et al., \prl \textbf{91}, 100403 (2003).
\end{thebibliography}
\end{document}